\documentclass[prl,twocolumn]{revtex4}%
\usepackage{amssymb}
\usepackage{amsfonts}
\usepackage{amsmath}
\usepackage{graphicx}%
\providecommand{\U}[1]{\protect\rule{.1in}{.1in}}

\begin{document}
\title{Why band theorists have been so successful in explaining and predicting novel superconductors?}
\author{I. I. Mazin}
\affiliation{Code 6393, Naval Research Laboratory, Washington, D.C. 20375, USA}
\date[Dated: ]{\today}

\begin{abstract}
In this contribution to the J. Phys. memorial issue in honor of Sandro
Massidda I reflect on a phenomenon Sandro had been a part of. While
theoretical condensed matter physicists have made, over the years, exciting
and most elegant contributions to the theory of superconductivity (which, in and
by itself, is one of the most beautiful constructs in theoretical physics),
some of them of utmost importance, they have had less success in predicting
and explaining superconducting states and mechanisms in specific materials.
More down-to-earth computational materials scientists, who often go by the
moniker \textquotedblleft band theorists\textquotedblright, have been much
more successful in applying (usually other people's) ideas in such
circumstances. In this essay I give some examples, largely drawn from my own
experience, and speculate on their meaning.

\end{abstract}
\maketitle

The last several decades have witnessed impressive contributions of band
theorists, now more often referred to as computational materials scentists, in
the fields of conventional and unconventional superconductivity. While the
former is amazing in terms of the accuracy and predictive power of such first
principles calculations, it is less surprising than the latter, which is
justly perceived to be a domain of \textquotedblleft high-brow
science\textquotedblright, well beyond direct capability of first principles
calculations. In my opinion, these two successes have quite different reasons,
and I will dicuss them both in the following text.

The first principles theory of (at that time conventional) superconductivity
goes back to George Gaspari and Balazs Gyorrfy, who suggested the first,
albeit rather approximate, technique for computing the electron-phonon
coupling constant, $\lambda,$ from $ab$ $initio$ band structure
calculations\cite{GG}, and fundamental theoretical works by Phil Allen, Dierk
Rainer, Eugene Maksimov and many others, well summarized in Rainer's
review\cite{DR}. As a result, a foundation of first principles calculations of
the anisotropic Eliashberg functions was established (a special mention goes
to collaborations between Phil Allen, mostly an analytical theorist, and band
theorists, such as Bill Butler, Warren Pickett, and others). After
establishing the technique of applying Sternheimer's perturbation
theory\cite{S} to the linear response in periodic solids by Baroni and his
collaborators\cite{B}, the community witnessed rapid progress in calculating
the Eliashberg function and solving the Eliashberg equation, first isotropic,
and then anisotropic, from first principles. A parallel, very successful
development, also based on these linear response techniques, was
superconducting density functional theory\cite{EKU}. As these two directions
matured (Sandro was taking an active part in both developments), accurate
calculations of the critical temperatures of simple classical superconductors,
such as Nb or Mo, had become possible. The very fact that a quantity
exponentially dependent on parameters (at least, in the weak coupling regime)
was coming out basically right was perceived as curious, but not necessarily
very enlightening. The fact that superconductivity of the doped fullerenes was
clarified in 1991 on the very basic level, as being related to intramolecular
bond-length-changing phonon modes by three independent groups\cite{VZ,Sl,MRA},
two of which entirely composed of band theorists, was left unappreciated,
partially because it was later recognized that the full physical picture was
more complicated than that.

This perception would change definitively in 2001 after the discovery of a 40
K superconductivity in MgB$_{2}.$ Recognized gurus of strongly correlated
matter, who had gained fame through high-Tc cuprates (we will get to this
family later) proffered all sorts of theories to this effect, all of them, it
turned out, spectacularly off mark. At the same time band theorists were able
to identify both the essential physcs, aptly dubbed by Warren Pickett
\textquotedblleft doping of covalent bonds\textquotedblright\ (we will get
back to this concept later)\cite{wep}, and the correct structure of the order
parameter (two distinctly different gaps)\cite{Amy}. In both cases the insight
was based, in a profound and intimate level, on the very material-specific
aspects of the electronic bands, Fermi surfaces, and the calculated Eliashberg
function\cite{SM}. Lack of this detailed understanding of materials-specific
electronic properties is what prevented bright minds fed on model Hamiltonians
from uncovering the right physics. Their \textit{modus operandi} was
\textquotedblleft what exciting physics is possible?\textquotedblright\ rather
than \textquotedblleft what is possible in this specific
material?\textquotedblright\ (I will come back to this important difference
later again)

It is instructive at this point to go back twelve years and look at the paper
by Len Mattheiss et al\cite{Len}. This was, arguably, the first paper where a
new superconducting compound was suggested theoretically and verified
experimentally. Noteworthy, this prediction was made years before the
computational soft- and hardware had developed to a level permitting full
electron-phonon coupling calculations for complex solids. Rather, the
prediction was based entirely on Len's band theorist's intuition. A few years
earlier he had calculated the electronic structure of the so-called Sleight
oxide, BaBiO$_{3},$ which had been known to superconduct at a few Kelvin upon
doping Pb for Bi, despite having a very low density of states. Mattheiss found
that the conducting electrons are primarily derived from the Bi orbitals, and
correctly conjectured that doping the \textquotedblleft
active\textquotedblright\ bands does not allow for full benefits for
superconductivity. Thus, he proposed to substitute the fully ionized Ba for K,
and, indeed, the critical temperature for the optimal doping was dramatically
increased compared to that of Ba(Bi,Pb)O$_{3},$ eventually surpassing 30 K.

This example shows that the main advantage of computational material
scientists over the model theories adepts is not, or, at least, mostly not,
the access to accurate numbers cranked by a computer, but material-sensitive,
chemistry-driven intuition developed through performing calculation and
analyzing their results for many classes of materials.

With time, more and more accurate calculations for more and more complex
systems have become possible, together with considerable progress in computing
crystallographic stability, eventually leading to excellent \textquotedblleft
brute force\textquotedblright\ predictions of entirely new superconductors.
This development has recently culminated in two record-breaking discoveries,
superconductivity in H$_{3}$S at 200 and in LaH$_{10}$ at 250 K, both
predicted theoretically as materials stable at high pressure and as
high-temperature superconductors, and, in fact, both utilizing Pickett's
concept of \textquotedblleft doped covalent bonds\textquotedblright.

This is probably less surprising given that the theory of phonon-driven
superconductivity is well established (as long as the key conditions are
satisfied, namely, $kT_{c}\sim\Delta\lesssim\omega_{ph}\ll E_{F}),$ and
materials in question are satisfactorily described by the density dunctional
theory, DFT, (which is, by nature, a static mean field theory). What is
astonishing is that band theorists (if not the band theory directly) have
provided over the years a lot of insight into the physics of unconventional
superconductivity in unconventional materials.

Historically, when it was first realized that the insulating state of the
parent compounds of the high-Tc cuprates cannot be reproduced by DFT
calculations, the first reaction was to deny the latter any relevance and
utility. To lesser extent, similar fate was suffered by Eliashberg equations.
This author was a part of a paper submitted originally to Phys. Rev.
Lett.\cite{ZZ}, which was rejected on the ground of its using the
Zeyher-Zwicknagl theory\cite{ZZZ}, dismissed by the referee as
\textquotedblleft uncritical utilization of the Eliashberg equations beyond
their range of applicability\textquotedblright, and calculating the
electron-phonon coupling parameters from DFT, \textquotedblleft application of
DFT to materials to which it has no relevance\textquotedblright. The excellent
agreement with the experiment did not seem to waver the referee's conviction.
Needless to say, DFT calculations of the electron-phonon coupling in optimally
doped cuprates were later recognized as quite reasonable, and regardless of
the nature of the superconducting gap, its effect on the phonon energy was
correctly described by the Zeyher-Zwicknagl formalism.

It is instructive to recall the state of experimental affairs in the field of
superconducting cuprates at that time. Both the ARPES techique and the sample
quality would undergo dramatic progress in the coming decade, but at that time
it was not even able to detect the Fermi surface in the optimally doped
cuprates. Superconductivity in alleged absence of a regular Fermi surface in
the normal state led to a plethora of highly exotic theories, including such
notions as, for instance, \textquotedblleft pseudo Fermi
surface\textquotedblright\ or anyon superconductivity. An idea that, at least
in the optimally- and overdoped high-Tc superconductors, the concept of
(renormalized) Fermi liquid may not be so out-of-touch with reality and the
DFT Fermi surface may actually be meaninful was, at that time, advocated only
by band theorists\cite{me}. As we now know very well, this concept was
correct, and improved experiments have confirmed it. For a while, it seemed
impossible to distinguish spectroscopically the bonding-antibonding splitting
in bilayer cuprates, which led to a new generations of intriguing physical
theories, such as \textquotedblleft interlayer pair
tunneling\textquotedblright\cite{IPT}. Again, with time the experiment has
converged to the DFT results\cite{Warren}, not the other way around, as it
often happens with theories. Moreover, while there is still no consensus on
the theory of superconductivity in cuprates, arguably the most popular
direction takes the most intuitive, from the point of view of DFT, approach,
wherein the pairing interaction has magnetic origin and the Fermi surface
geometry and the character of spin fluctuations (both easily obtainable from
DFT calculations) lead to a d-wave pairing. This point of view was advocated
by Doug Scalapino, David Pines and others. Of course, they were basing their
intuition on the experimental findings, but could have used DFT calculations
as well.

This said, it is quite clear that the essential physics of the high-Tc
superconductivity is beyond the scope of DFT, which brings us to another
important aspect of the latter. DFT is a \textit{quantitative }theory, based
on a well-defined set of approximations. It is in a unique position where both
successes and failures of a theory add equally to our understanding of real
physics in a particular system. Model calculations more often than not can be
tuned (to avoid the word \textquotedblleft massaged\textquotedblright) to
agree with essentially any experiment. DFT has relatively few knobs (although
its extentions such as DFT+U or DFT+DMFT have more), and therefore its failure
clearly outlines the applicability of its fundamental assumptions to a given
material. Moreover, depending on the manner in which it fails one can extract
invaluable information about the system. For instance, the fact that DFT
underestimates the tendency to magnetism in cuprates tells us about the
importance of local physicis and local correlations there. Conversely, the
fact that DFT \textit{overestimates }the tendency to magnetism in Fe-based
superconductors suggests that magnetism there is largely itinerant and
suppressed by long-range spin fluctuations.

Let us continue our historical excursion. For decades, Cu-based
superconductors occupied most of researchers' attention, however, there were
interesting splashes time and again in the sea of superconducting materials.
The first that comes to mind is \textit{A}$_{3}$C$_{60}$ ($A=$K,Rb,Cs).
Superconductivity at a record at that time (not counting the cuprates)
temperatures up to 40 K\cite{K3C60} was mentioned above. It is worth noting
that it was not a brute force calculation that provided insight; in fact, full
electron-phonon calculations as we know them now were not possible at that
time, and when they became possible it turned out that they underestimate the
calculated coupling constant by 40\% (band calculations beyond DFT, such as
hybrid or GW functional correct for this error). Rather, it was qualitative
(and correct) understanding of the basic physics, which had emerged from
relatively inaccurate DFT calculations.

Fast forward ten years, and we have a bunch of discoveries that would have
been sensational before the cuprates, but only attracted moderate attention in
2001. I am speaking of several new superconductors discovered, or declared in
that year: MgCNi$_{3},$ ZrZn$_{2},$ $\varepsilon$-Fe, and, of course,
MgB$_{2}.$ Among them we find, in this order, (i) a material with a very strong
electron-phonon coupling, but with a rather modest critical temperature of 8
K\cite{MgCNi3}, (ii) an experimental artifact, a several years later retracted
claim of coexistence of superconductivity and ferromagnetism\cite{ZrZn2},
(iii) a ferromagnet turned into a superconductor by pressure\cite{e-Fe}, and
(iv) the record normal-pressure-Tc conventional superconductor\cite{MgB2}. In
all these cases the original assessement of the community was to a larger or
smaller extent off the mark. In the first case, the original paper was
pointing out an analogy with the LnNi$_{2}$B$_{2}$C family, which are 2D
superconductors, and expressed a justified surprise that in a 3D analogue
$T_{c}$ is much lower. This paper  advocated for an unconventional
superconductivity, having no gauge for the strength of the electron-phonon
coupling. Early DFT calculations, however, have identified a very strong
coupling with rotational phonon modes\cite{MgCNi3us} (with, it was later
found, a strong anharmonic component\cite{SS}), with a strong pair-breaking
due to proximity to magnetism.

In the second case, the original paper was aiming at triplet pairing,
postulating a microscopic coexistance with ferromagnetism; DFT calculations,
again, showed strong coupling with particular phonons, and the calculated
electronic and magnetic properties did not look encouraging for triplet
pairing --- nothing explicit, just a general impression. Correspondingly, Ref.
\cite{ZrZh2us} offered three possible scenarios, ordered by their likelihood:
sample inhomogenuity, a Fulde-Ferrel state, or, least likely, triplet pairing.
Indeed, a few years later it was found\cite{ZrZn2true} that Ref. \cite{ZrZn2}
had observed surface superconductivity.

Finally, in the third case it was found that when Fe loses its ferromagnetism
at pressures above 15 GPa it transforms into a hexagonal structure know as
$\varepsilon$-Fe and becomes superconducting. The original paper was calling
$\varepsilon$-Fe \textquotedblleft non-magnetic\textquotedblright\ and
referred to decades-old predictions (including one involving Sandro
Massidda)\cite{Fe} that the electron-phonon coupling in Fe, \textit{if the
magnetism could be entirely suppressed}, would be strong enough to provide for
measureable superconductivity. When band theorists took a closer look at this
experiment, they found that (1) $\varepsilon$-Fe at this pressures is not
\textit{non-magnetic}, but \textit{paramagnetic}, sporting large, but
disordered and strongly fluctuating magnetic moments\cite{Fe-Ron} and (2)
electron-phonon coupling depends on the pressure very weakly, while
superconductivity exists only in a very narrow pressure range\cite{Fe-us}.
These two finding were in contradiction with the original proposal of the
phonon-driven superconductivity, but rather pointing toward a
spin-fluctuations-induced, uncoventional pairing. Subsequent experiments found
unusual sensitivity to disorder, indicating such an unconventional
pairing\cite{Jackard}.

Let us once again return to the fourth case, MgB$_{2},$ which was, arguably,
the most important discovery in superconductivity since the cuprates. The
computational part, even while challenging at that time, was rather trivial,
as was the much touted prediction of a two-gap superconductivity\cite{Amy}
(curiously, the \textquotedblleft model Hamiltonian\textquotedblright%
\ theorists were quite at loss with this compound, apparently still under the
spell of the incorrect claim of a 30 K $T_{c}$ limit for conventional
superconductors, see Ref.\cite{VL}; they were proposing such exotic mechanisms
as acoustic phonons, resonant valence bond, etc., overlooking the simple
solution). What was not trivial was a new concept in phonon-driven
superconductivity, introduced by Warren Pickett\cite{wep}, which has proven
instrumental for recently dicovered record-Tc hydorgen based superconductors:
doped covalent bonds.

An opposite idea, of the utility of soft modes, had been circulated in the
community for decades, even though many theorists were warning that the
classical soft modes are only soft at particular points in the Brillouin zone
and thus do not have enough phase space to boost superconductivity. The soft
mode concent was based on the mathematically correct statement that phonons
with frequencies close to $2\pi T_{c}$ are the most efficient in increasing
the critical temperature\cite{RMP}, and that the classical expression for the
electron-phonon coupling includes inverse force constants, which can written
as $\lambda_{\nu,\mathbf{q}}\varpropto1/\Phi=1/M\omega_{\nu,\mathbf{q}}^{2},$
where $M$ is some effective ionic mass. Obviously, lowering $\omega$ will
increase $\lambda$ in the exponent, with should be more important than the
prefactor $\omega.$ Or is it?

Pickett's observation related to the dual character of the electron-phonon
coupling: while indeed $\lambda\approx\eta/\Phi,$ where $\eta$ is an
electronic factor characterizing the sensitivity of the electronic structure
to ionic displacements, and $\Phi$ is some measure of the force constants, the
force constants themselves depend on $\eta,$ roughly as $\Phi\approx\Phi
_{0}-2\eta,$ where $\Phi_{0}$ is the unscreened force constants parameter.
Thus, $\eta$ cannot be too large, because of the danger of the lattice
becoming unstable. Doping covalent bonds is difficult to effect, but, if it
can be done, it achieves two goals. First, the electronic structure strongly
depends on ionic displacements modulating the covalent bond lengths (such are
modes responsible for high $T_{c}$ in both $A_{3}$C$_{60}$ and in MgB$_{2}),$
providing for large $\eta,$ and the unscreen force constants are hard for
covalent bonds, that providing for a large $\Phi_{0}.$ Taken together, one can
have a large $\lambda$ $and$ a large $\omega.$ This is exactly what happens in
MgB$_{2}.$ The $E_{g}$ modes, mainly responsible for superconductivity, have
large $\Phi_{0},$ of the order of 10$^{7}$ a.m.u$\times$cm$^{-2},$ which is
reduced by a factor of three by the electron-phonon coupling --- and still the
phonon frequencies remain high, on the order of $500$ cm$^{-1}.$ The same
physics appears to be crucial for the recently discovered record-temperature
superconductors H$_{3}$S\cite{H3S} and LaH$_{10}$\cite{LaH10} (which,
incidentally, were both predicted computationally). A novel element there is
that ultra-high pressures were instrumental in doping the H-involving covalent
bonds, simply because compression severly limits possibilities of a system
(which is obviously unhappy to have metallic covalent bonds) to assume a
different crystal structure and get read of this undesired property.

Between 2001 and 2018 a number of interesting novel superconductors were
discovered, and in explaining most, if not all of them, band theorists played
an important and often a leading part. I will not dwell on those, having shown
enough very diversified examples already. One milestone is however hard to
pass over: Fe-based superconductors. These are the only truly high-Tc
unconventional superconductors besides cuprates. The history of MgB$_{2}$ had
repeated itself again, in the sense that outstanding theorists were suggesting
all sort of highly exciting scenarios, but it was band theorists who came up
with the basic idea (now understood to be a big oversimplification, but still
covering the basic physics) of a novel pairing state dubbed $s_{\pm}$ since
then\cite{FeBS}. There was a simple reason for that. This novel
superconducting state basically results from a fortuitous combination of a
particular Fermi surface geometry and a particular type of magnetic ordering
these systems are close to. Both things are something band theorists calculate
routinely and have a good intuition about.

Which brings to the crucial question: what is the advantage that band
therists, a.k.a. computational materials scientist, have? The author of this
essay is of firm opinion that theorists doing model calculations are generally
smarter than us, and have deeper knowledge of the theory of superconductivity.
Despite that, their intuition as regards specific superconducting materials
seems to work less well. A seemingly unrelated, but in reality quite similar
question is, why solid state chemists, and not physicists, are responsible for
a lion share of novel superconductors discovered in the last several decades?
Materials-dependent intuition may be the key. Arguably, the most interesting
question a theoretical physicst can ask is, what can happen, in principle?
Answers to this question often lead to most interesting toeretical models,
exciting possibilities, new concepts. An alternative question to ask is, what
can happen \textit{in this specific material}? Generically, this question is
more boring, but also more productive as regards understanding a concrete
superconductor. A correct answer to this question requires both a profound
understanding of the microscopic physics, specific features of electronic
bands, phonon spectra, magnetic orders, and an intuitive feeling based on
dozen, if not handreds previous calculations indicating what is possible in
which material. Last but not least, in order to predict something one need not
necessarily invent something new, but rather be aware of all hypotetical idea
that had been floated around, sometimes decades ago, by smarter-than-us
theorists. It is worth noting that the two gap superconductivity had been
studied, as a purely theoretical concept, as early as the late 50s\cite{multi}%
; the $s_{\pm}$ superconductivity in the early 70s.\cite{spm} Without knowing
this general (and, until some time, purely abstract) frameworks it would be
hard to associate actual superconductivity in MgB$_{2}$ and in LaFeAsO with
these phenomena.

In conclusion, I would like to express my gratitude to my many
brothers-in-arms, who over the years worked in the same field, applying band
structure calculations to supercondutivity, as Sandro, for instance, did. I am
also thankful to my senior colleagues, my teachers and role models: Ole
Andersen, Warren Pickett, Phil Allen, as well as to those who initiated me
into the theory of superconductivity: Eugene Maksimov, David Kirzhnits, Vitaly Ginzburg.


\begin{thebibliography}{99}
\bibitem {SM}I cannot resist but mentioning that Sandro Massidda has made an
invaluable contribution into study of MgB$_{2},$ having published 30 papers on
this subject, with more than 1000 cummulative citation count.
\end{thebibliography}
\end{document}